# Data Quality Measures and Data Cleansing for Research Information Systems

Otmane Azeroual[1, 2, 3], Gunter Saake[2], Mohammad Abuosba[3]
[1] German Center For Higher Education Research And Science Studies (Dzhw), 10117 Berlin, Germany
[2] Otto-Von-Guericke University Magdeburg, Department Of Computer Science Institute For Technical And Business Information Systems Database Research Group P.O. Box 4120; 39106 Magdeburg, Germany
[3] University Of Applied Sciences Htw Berlin Study Program Computational Science And Engineering Wilhelminenhofstrasse 75a, 12459 Berlin, Germany

Otmane Azeroual: Azeroual@dzhw.eu
Gunter Saake: Saake@iti.cs.uni-magdeburg.de
Mohammad Abuosba: Mohammad.Abuosba@HTW-berlin.de

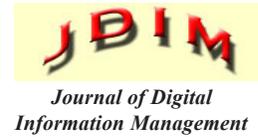

*Journal of Digital Information Management*

**ABSTRACT:** *The collection, transfer and integration of research information into different research information systems can result in different data errors that can have a variety of negative effects on data quality. In order to detect errors at an early stage and treat them efficiently, it is necessary to determine the clean-up measures and the new techniques of data cleansing for quality improvement in research institutions. Thereby an adequate and reliable basis for decision-making using an RIS is provided, and confidence in a given dataset increased.*

*In this paper, possible measures and the new techniques of data cleansing for improving and increasing the data quality in research information systems will be presented and how these are to be applied to the research information.*

## 1. Introduction

With the introduction of research information systems, research institutions can have a current overview of their research activities, record, process and manage information about their research activities, projects and publications as well as integrate them into their website.

For researchers and other interest groups, they offer opportunities to collect, categorize and use research information, be it for publication lists, preparing projects, reducing the burden of producing reports or outlining their research output and scientific expertise.

To introduce a research information system means for research institutions to provide their required information on research activities and research results in a secured quality. Because the problems of poor data quality can spread across different domains and thereby weaken entire research activities of an institution, the question arises of what measures and new techniques can be taken to eliminate the sources of error, to correct the data errors and to remedy the causes of errors in the future.

In order to improve and increase the data quality in research information systems, this paper aims will be to present the possible measures and new techniques of data cleansing which can be applied to research information.

## 2. Research Information and Research Information Systems

In addition to teaching, research is one of the core tasks of universities. Information about task fulfillment and services in this area must be available with less time and more reliably. For the research area, data and information are mainly collected to map the research activities and their results, and to administer the processes associated with the research activity. This may include information on research projects, their duration, participating researchers and related publications. This information is also called research information or research data. The categories of research information and their relationships can be structured and summarized using Figure 1 of five dimensions:

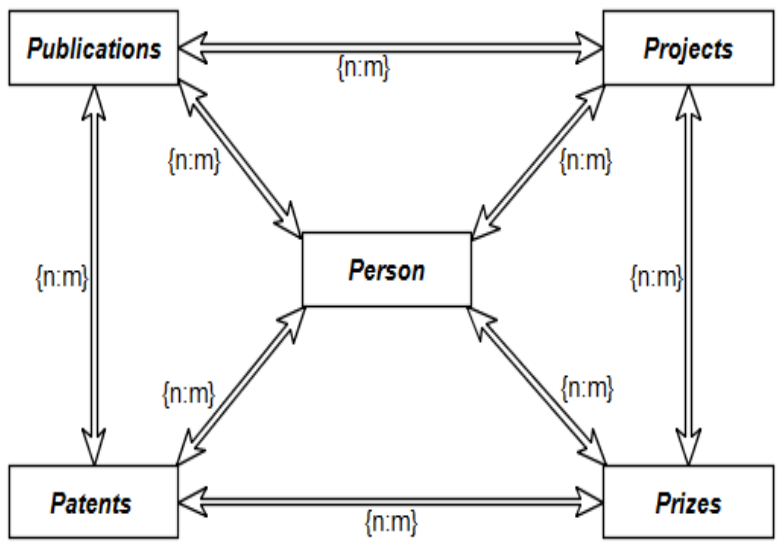

Figure 1. Dimensions of Research Information (RI)

As shown in the illustration 1, persons, as a scientist, are researching, doing their PhD or habilitation stand in the centre at various research institutions. In addition to projects, individuals can increase the scientific reputation of the publications they have written, as well as of prizes or patents received. Therefore, the relationship between this information represents an N-M relationship.

Processes and systems that store and manage the research information correspond to Research Information Systems (RIS or CRIS for Current Research Information System).

A RIS is a **central database** that can be used to collect, manage and deliver information about research activities and research results.

The following figure (see Figure 2) shows an overview of the integration of research information from a university into the research information system:

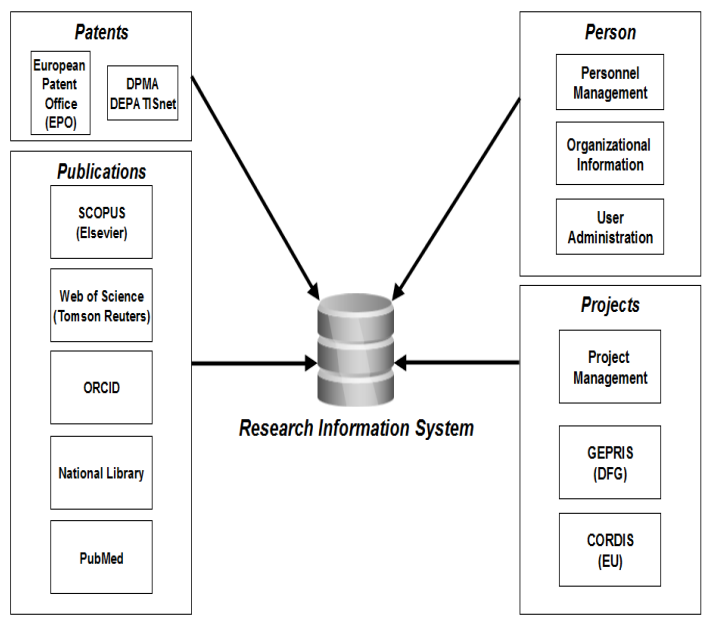

Figure 2. Integration of Research Information in the Research Information System (RIS)

In the RIS, data from a wide variety of existing operational systems are integrated in a research information system and are made by business intelligence tools in the form of reports available. Basically, all data areas of the facilities are mapped in the RIS; Filling takes place via classic ETL (Extract, Transform, Load) processes from both SAP and non-SAP systems.

RIS provide a holistic view of research activities and results at a research institution. They map the research activities not only institutions but also researchers current, central and clear. By the central figure in the system a working relief is possible for the researchers. Data is entered once with the RIS and can be used multiple times, e.g. on websites, for project applications or reporting processes. A double data management and with it an additional work for the users should be avoided. Another objective is to establish research information systems as a central instrument for the consistent and continuous communication and documentation of the diverse research activities and results. Improved information retrieval helps researchers looking for collaborators, companies in the allocation of research contracts, to provide the public with transparency and general information about their institution.

**3. Data Quality and its Problem in the RIS**

When interpreting the data into meaningful information, data quality plays an important role. Almost all users of the RIS recognized the importance of data quality for electronically stored data in a database and in research information systems.

The term data quality is understood to mean a "multidimensional measure of the suitability of data to fulfill the purpose of its acquisition/generation. This suitability may change over time as needs change" [21]. This underscores the subjective demands on data quality in each facility and illustrates a potential dynamic data quality process. The definition makes it clear that "the quality of data depends on the time of consideration and the level of demands made on the data at that time" [1].

To better understand the enormous importance of data quality, it is important to understand the causes of poor quality data that can be found in research information systems. Figure 3 below illustrates these typical causes of data quality deficiencies in data collection, data transfer, and data integration from data sources to the RIS.

| Processes of the RIS | Data Quality Defects |
|---|---|
| Data Collection | Incorrect data caused by input errors (Typos, Missing Information, Contradictory Information, Redundant Data Collection, Obsolete Data Attributes, Incomplete Data and Irrelevant Data Attributes) |
| Data Transfer | Technical errors in the transfer of data from the data sources to the RIS (e.g. in the form of faulty data carriers), Incorrect data processing processes for preparation or follow-up of the transmission (for example export from a database) |
| Data Integration | Incorrect transformation and cleansing processes of unification and consolidation of data |

Figure 3. Typical Causes of Data Quality Defects

Such quality problems in the RIS must be analyzed and then remedied by data transformation and data cleaning. Figure 4 below shows a practical example of these typical quality problems of data in the context of a RIS.

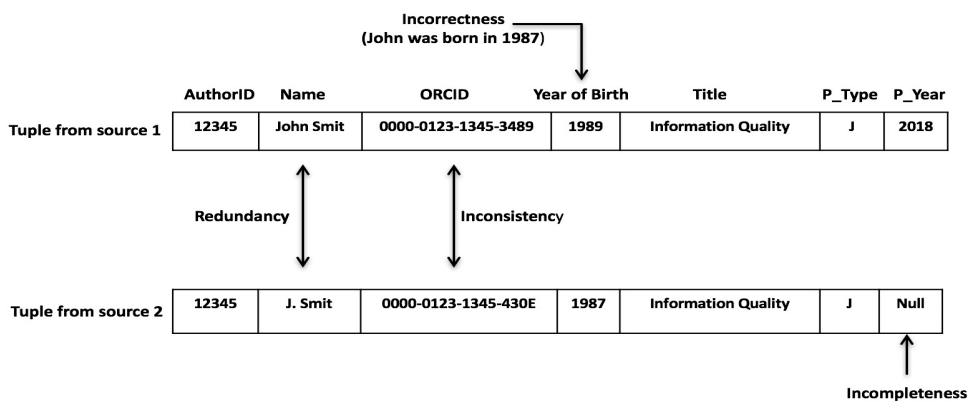

Figure 4. Examples of Data Quality Problems in the RIS

## 4. Data Quality Measures

In order to enable the most efficient process for detecting, improving and increasing data quality in RIS [22]:

• To analyze specifically the resulting data errors,

• To define what action needs to be taken to eliminate the sources of error and to clean up the erroneous data and

• Prioritize the various data errors and the necessary measures and transfer them to a corresponding project plan.

Only if these steps were carried out conscientiously, a free from problems and focused data cleansing can take place. The procedure for securing and increasing the quality of data in RIS is considered here after three measures [22] and the choice of the optimal procedure depends on the frequency of changes of the data and their importance for the user, as shown in Figure 5:

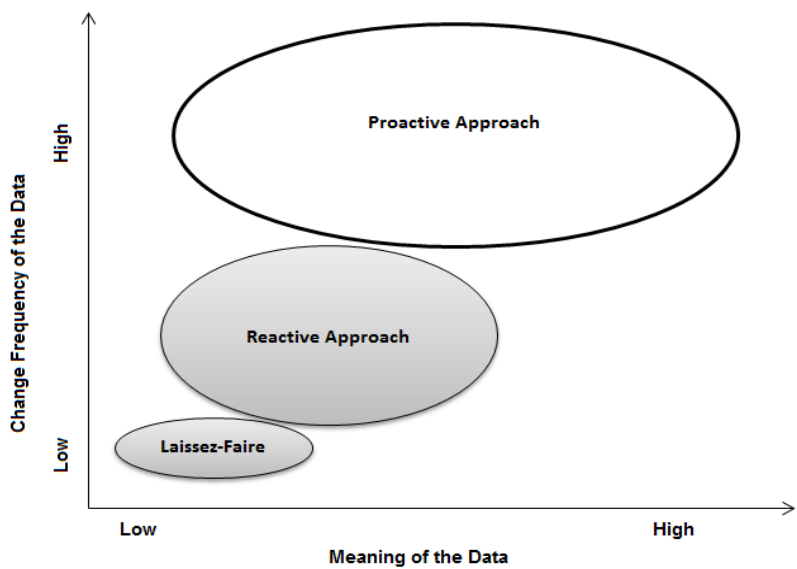

Figure 5. Measures Portfolio According to Different Criteria [19]

**Laissez-Faire**
For less and rarely changing data, the laissez-faire principle is appropriate. In this measure, the errors occurring without treatment are accepted. In other words, they are simply ignored or, if so, only incidentally resolved so as not to stop the business process.

**Reactive Approach**
For important and only rarely changing data, the reactive measures are suitable. They start directly with the data and correct quality defects, without, however, eliminating their cause. This cleanup can be done manually or by machine. Here no longer monitoring of the data quality takes place and the measures are always taken acutely and selectively [22].

**Proactive Approach**
On the other hand, proactive or (preventive) measures are available for important and frequently changing data. Here, measures are mainly taken to eliminate the sources of error and to prevent the occurrence of the errors. Continuous monitoring of possible errors takes place, as well as continuous measures to eliminate and prevent them.

The importance of the data is determined by the specific business processes of a facility. For example, this can be quantified by the sum of costs that are caused by possible quality deficiencies in these data. Depending on their nature, data errors can lead, among other things, to image damages, additional expenditures or wrong decisions.

It can be deduced from the change frequency of the data how long it takes for data to reach the lacking initial data quality level after an initial cleanup.

Figure 5 shows that the more frequently data changes, the more rewarding a proactive approach. The accumulated costs for one-off adjustments increase with the frequency of data changes. The more frequently they have to be applied, the lower the time saved and the cost involved compared to a proactive approach. According to [8], causes of poor data quality can be identified and adequate measures for quality improvement or increase can be identified. The identification of the causes of poor data quality is made possible by the "consideration of the entire process of data creation up to the data use with all related activities regarding qualitative objectives" [8].

Cleanup measures are only conditionally used to improve incomplete, missing, incorrect, or inconsistent data. The decision to use a particular measure must be made by a domain expert who is familiar with the business processes of his facility and can evaluate how quality deficiencies would adversely affect them [6] [19].

A continuous analysis of the quality requirements and a periodical control are necessary to ensure a high data quality in the long term. The following is the process for evaluating data quality. This requires consideration of the definition of data quality and the techniques presented to detect erroneous data. Therefore, the research institutions evaluate their data quality on the basis of different dimen- sions (such as completeness, timeliness, correctness and consistency), which correspond to their furnishing context, their requirements or their risk level [5].

The following model in Figure 6 contains a differentiated view and different weighting of quality dimensions and strives for a lasting assurance of data quality. The application of such a method can form the basis for an agile, effective and sustainable data quality control in research information systems.

The procedure for assessing the quality of data is described in six or seven steps [5]:

1. Identification of the data to be checked for data quality (typically research-related operational data or data in decision-making reports).

2. Deciding which dimensions are used and their weighting.

3. Work out an example of good and bad data for each criterion (several conditions can play a role here).

4. Apply the test criterion to the data.

5. Evaluate the test result and whether the quality is acceptable or not.

6. Make corrections where necessary (for example, clean up, improve data handling process to avoid repeated errors in the future).

7. Repeat the process periodically and observe trends in data quality.

**5. Data Cleansing**

The process of identifying and correcting errors and inconsistencies with the aim of increasing the quality of given data sources in the RIS is referred to as "Data Cleaning" or "Data Cleansing".

Data Cleaning includes all the necessary activities to clean dirty data (incomplete, incorrect, not currently, inconsistency, redundant, etc.). The data cleansing process can be roughly structured as follows [9]:

1. Define and determine the actual problem.

2. Find and identify faulty instances.

3. Correction of errors found.

Data Cleansing uses a variety of special methods and technologies within the data cleansing process. This is divided into five areas (see Figure 7):

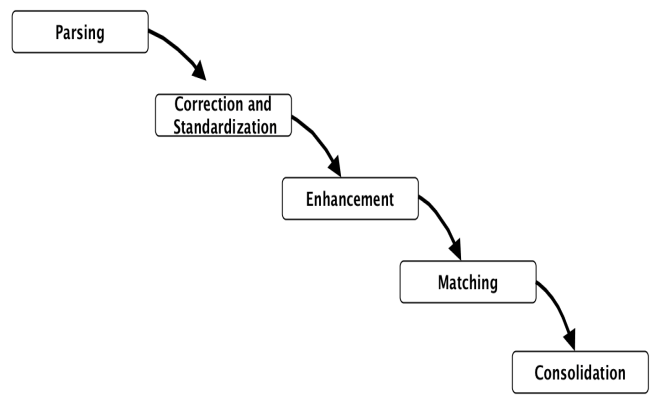

Figure 7. Data Cleansing Process

**Parsing** is the first critical component of data cleansing and helps the user understand and transform the attributes. Individual data elements are referenced according to the metadata. This process locates, identifies and isolates individual data elements. For this process, e.g. for names, addresses, zip code and city. The parser Profiler analyzes the attributes and generates a list of tokens from them, and with these tokens the input data can be analyzed to create application-specific rules. The biggest problem here are different field formats, which must be recognized.

**Correction & Standardization** is further necessary to check the parsed data for correctness, then to subsequently standardize it. Standardization is the prerequisite for successful matching and there is no way around using a second reliable data source. For address data, a postal validation is recommended.

**Enhancement** is the process that expands existing data with data from other sources. Here, additional data is added to close existing information gaps. Typical enrichment values are demographic, geographic or address information.

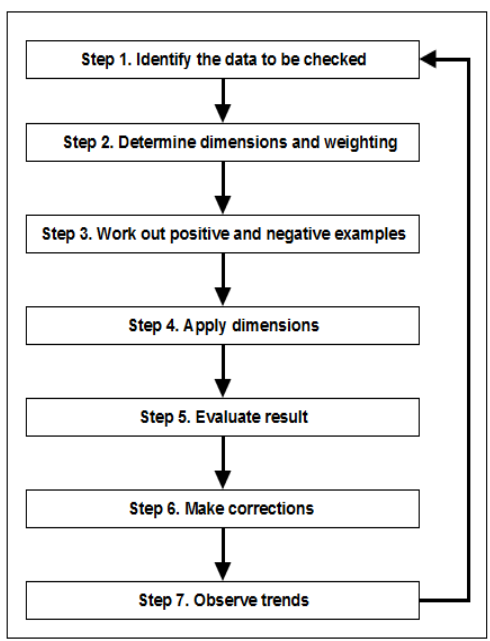

Figure 6. Procedure Model for Data Quality Assessment [5]

**Matching** There are different types of matching: for reduplicating, matching to different datasets, consolidating or grouping. The adaptation enables the recognition of the same data. For example, redundancies can be detected and condensed for further information.

**Consolidation (Merge)** Matching data items with contexts are recognized by bringing them together.

All of these steps are essential for achieving and maintaining maximum data quality in research information systems. Errors in the integration of multiple data sources in a RIS are eliminated by the cleanup.

The following Table 1 illustrates an example of identifying records with faulty names in a publication list to show how data cleansing processes (cleanup, standardization, enrichment, matching, and merging) can improve the quality of data sources.

The cleanup process adds missing entries, and completed fields are automatically adjusted to a specific format according to set rules.

**Original Data before cleanup**

| Data Source | | | | |
|---|---|---|---|---|
| Author ID | Name | ORCID | Birth Date | Address |
| 12345 | John Smit | 0000-0123-1345-3487 | 12/23/1987 | 123 6th Street, Melbourne, 32904 |
| 12345 | Dr. John Smit | 0000-0000-0000-0000 | 23.12. 1987 | 6th Street, 32904 123 |
| 12345 | John William Smit | 0000-0123-1345-3487 | 872312 | 10 Street 32904 6th |
| 14587 | J. Smit | | 09/23/78 | 71 Pilgrim Ave. 32904 |
| 14587 | Smit John | 0000-0123-1345-3487 | 23.09.1987 | |
| | John Smit | 0000012313453487 | 23.12.1987 | Street, 32904 6th US |
| 19875 | Lena Scott | 0001-0254-4118-F006 | 14-1-1984 | 44 Shirley Ave. West Chicago 60185 |
| 19875 | Scott Lena | 000102544118F006 | 1984 | Shirley Ave. West Chicago 60185 |

Table 1. Example of missing entries

**Data after Cleanup**
In this example, the missing zip code is determined based on the addresses and added as a separate field. Enrichment rounds off the content by comparing the information to external content, such as demographic and geographic factors, and dynamically expanding and optimizing it with attributes.

| Cleansed Data | | | | | |
|---|---|---|---|---|---|
| Author ID | First | Last | ORCID | Birth Date | Address |
| 12345 | John | Smit | 0000-0123-1345-3487 | 1987-12-23 | 32904; FL; Melbourne; 123 6th Street |
| 12345 | John | Smit | | 1987-12-23 | 32904; FL; Melbourne; 123 6th Street |
| 12345 | John | Smit | 0000-0123-1345-3487 | | 32904; FL; Melbourne; 123 6th Street |
| 14587 | John | Smit | | 1978-09-23 | 32904; FL; Melbourne; 71 Pilgrim Ave. |
| 14587 | John | Smit | 0000-0123-1345-3487 | 1987-09-23 | |
| | John | Smit | | 1987-12-23 | 32904; FL; Melbourne; 123 6th Street |
| 19875 | Lena | Scott | 0001-0254-4118-F006 | 1984-01-14 | 60185; IL; West Chicago; 44 Shirley Ave. |
| 19875 | Lena | Scott | 0001-0254-4118-F006 | | 60185; IL; West Chicago; 44 Shirley Ave. |

**Data before Enrichment**

| Cleansed Data | | | | | |
|---|---|---|---|---|---|
| Author ID | First | Last | ORCID | Birth Date | Address |
| 12345 | John | Smit | 0000-0123-1345-3487 | 1987-12-23 | FL; Melbourne; 123 6th Street |
| 12345 | John | Smit | | 1987-12-23 | FL; Melbourne; 123 6th Street |
| 12345 | John | Smit | 0000-0123-1345-3487 | | FL; Melbourne; 123 6th Street |
| 14587 | John | Smit | | 1978-09-23 | FL; Melbourne; 71 Pilgrim Ave. |
| 14587 | John | Smit | 0000-0123-1345-3487 | 1987-09-23 | |
| | John | Smit | | 1987-12-23 | FL; Melbourne; 123 6th Street |
| 19875 | Lena | Scott | 0001-0254-4118-F006 | 1984-01-14 | IL; West Chicago; 44 Shirley Ave. |
| 19875 | Lena | Scott | 0001-0254-4118-F006 | | IL; West Chicago; 44 Shirley Ave. |

**Data after Enrichment**
The example shows how the reconciliation and merge process runs. Merging and matching promote consistency because related entries within a system or across systems can be automatically recognized and then linked, tuned, or merged.

| Enriched Data | | | | | | |
|---|---|---|---|---|---|---|
| Author ID | First | Last | ORCID | Birth Date | Address | Zip |
| 12345 | John | Smit | 0000-0123-1345-3487 | 1987-12-23 | FL; Melbourne; 123 6th Street | 32904 |
| 12345 | John | Smit | | 1987-12-23 | FL; Melbourne; 123 6th Street | 32904 |
| 12345 | John | Smit | 0000-0123-1345-3487 | | FL; Melbourne; 123 6th Street | 32904 |
| 14587 | John | Smit | | 1978-09-23 | FL; Melbourne; 71 Pilgrim Ave. | 32904 |
| 14587 | John | Smit | 0000-0123-1345-3487 | 1987-09-23 | | |
| | John | Smit | | 1987-12-23 | FL; Melbourne; 123 6th Street | 32904 |
| 19875 | Lena | Scott | 0001-0254-4118-F006 | 1984-01-14 | IL; West Chicago; 44 Shirley Ave. | 60185 |
| 19875 | Lena | Scott | 0001-0254-4118-F006 | | IL; West Chicago; 44 Shirley Ave. | 60185 |

**Matching**

This example finds related entries for John Smit and Lena Scott. Despite the similarities between the datasets, not all information is redundant. The adjustment functions evaluate the data in the individual records in detail and determine which ones are redundant and which ones are independent.

| Cleansed Data | | | | | |
|---|---|---|---|---|---|
| Author ID | First | Last | ORCID | Birth Date | Address |
| 12345 | John | Smit | 0000-0123-1345-3487 | 1987-12-23 | 32904; FL; Melbourne; 123 6th Street |
| 12345 | John | Smit | | 1987-12-23 | 32904; FL; Melbourne; 123 6th Street |
| 12345 | John | Smit | 0000-0123-1345-3487 | | 32904; FL; Melbourne; 123 6th Street |
| 14587 | John | Smit | | 1978-09-23 | 32904; FL; Melbourne; 71 Pilgrim Ave. |
| 14587 | John | Smit | 0000-0123-1345-3487 | 1987-09-23 | |
| | John | Smit | | 1987-12-23 | 32904; FL; Melbourne; 123 6th Street |
| 19875 | Lena | Scott | 0001-0254-4118-F006 | 1984-01-14 | 60185; IL; West Chicago; 44 Shirley Ave. |
| 19875 | Lena | Scott | 0001-0254-4118-F006 | | 60185; IL; West Chicago; 44 Shirley Ave. |

**Consolidation**

The merge makes the reconciled data a comprehensive data set. This example merges the duplicate entries for John Smit into a complete record that contains all the information.

| Cleansed Data | | | | | |
|---|---|---|---|---|---|
| Author ID | First | Last | ORCID | Birth Date | Address |
| 12345 | John | Smit | 0000-0123-1345-3487 | 1987-12-23 | 32904; FL; Melbourne; 123 6th Street |
| 12345 | John | Smit | | 1987-12-23 | 32904; FL; Melbourne; 123 6th Street |
| 12345 | John | Smit | 0000-0123-1345-3487 | | 32904; FL; Melbourne; 123 6th Street |
| 19875 | Lena | Scott | 0001-0254-4118-F006 | 1984-01-14 | 60185; IL; West Chicago; 44 Shirley Ave. |
| 19875 | Lena | Scott | 0001-0254-4118-F006 | | 60185; IL; West Chicago; 44 Shirley Ave. |

| Golden Record | | | | | |
|---|---|---|---|---|---|
| Author ID | First | Last | ORCID | Birth Date | Address |
| 12345 | John | Smit | 0000-0123-1345-3487 | 1987-12-23 | 32904; FL; Melbourne; 123 6th Street |
| 19875 | Lena | Scott | 0001-0254-4118-F006 | 1984-01-14 | 60185; IL; West Chicago; 44 Shirley Ave. |

## 6. Discussion

Cleaning up of data errors in RIS is one of the possible ways to improve and enhance existing data quality. According to the defined measures and data cleansing processes, the following developed use cases could be identified in the target system and serve as a model to show how to detect, quantify, correct, improve and increase them in case of data errors in research information systems in the facilities and research institutions.

The following figure introduces the just mentioned use case for improving and increasing the data quality in the RIS.

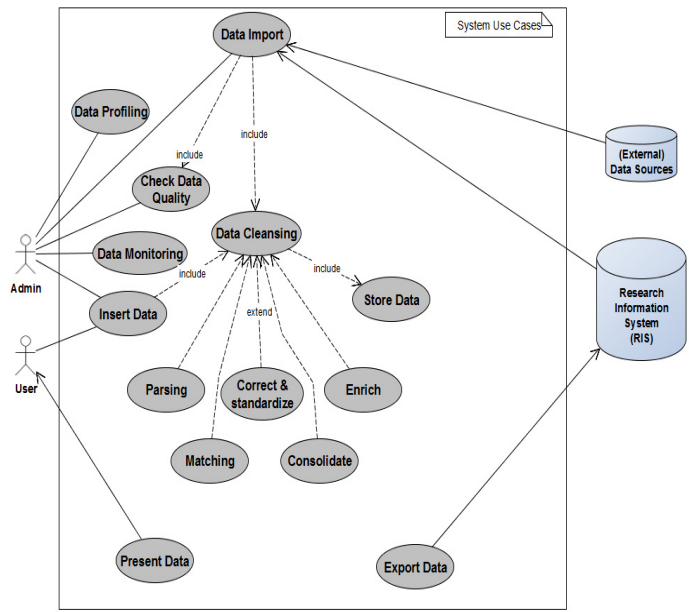

Figure 8. Use Case for improving and increasing the Quality of Data in the RIS

The meta process flow can be viewed as shown in the following figure.

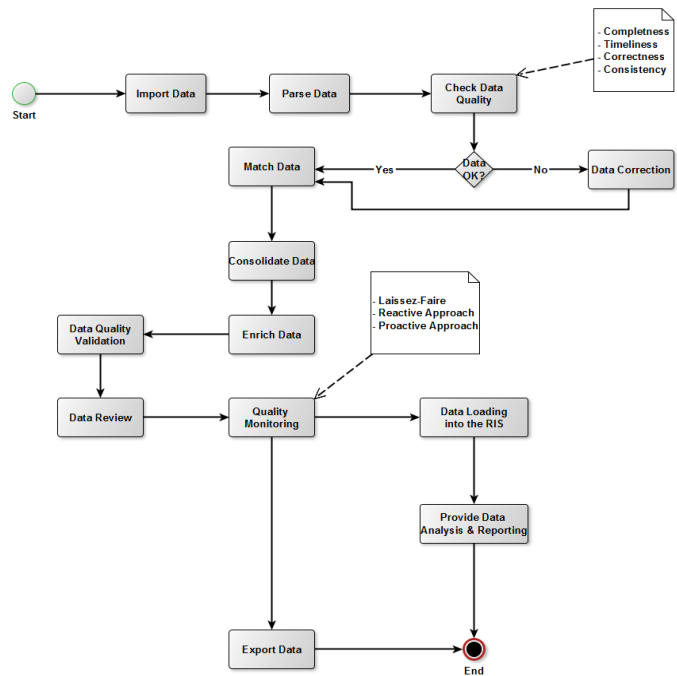

Figure 9. Main Workflow of the process

The proposed measures and the new techniques of data cleansing help institutions to overcome the identified problems. With these steps every facilities and research institutions can successfully enforce their data quality.

## 7. Conclusion and Outlook

In research, a direct positive correlation between data quality and RIS is predominantly discussed. An improvement in data quality leads to an improvement in the information basis for RIS users. In addition, this paper addresses the question of which measures and new techniques for improving and increasing data quality can be applied to research information. The aim was to present possible measures and the new techniques of data cleansing for improving and increasing the data quality in research information systems and how these should be applied to the research information.

After identifying an error, it is not enough to just clean it up in a mostly time-consuming process. In order to avoid reoccurrence, it should be the aim to identify the causes of these errors and to take appropriate measures and new techniques of data cleansing. The measures must first be determined and evaluated. As a result, it is clearly shown that a proactive approach is the safest method to guarantee high data quality in research information systems. However, depending on the nature of the failures, satisfactory results can be achieved even with the less expensive and extravagantly reactive or laissez-faire practices.

As a result, the new techniques of data cleansing demonstrate that the improvement and enhancement of data quality can be made at different stages of the data cleansing process for each RIS, and that high data quality allows universities and research institutions e.g. to operate RIS successfully. In this respect, the review, improvement and enhancement of data quality is always purposeful. The illustrated concept can be used as a basis for the using facilities. It offers a suitable procedure and a use case, on the one hand to be able to better evaluate the data quality in RIS, to prioritize problems better and to prevent them in the future as soon as they occur. On the other hand, these data errors have to be corrected, improved and increased with measures and data cleansing. It says, "The sooner quality problems are identified and remedied, the better." There are many tools to help universities and all research institutions perform the data cleansing process. With these tools, all facilities such as completeness, correctness, timeliness, and consistency of their key data can be significantly increased and they can successfully implement and enforce formal data quality policies.

Data Cleansing tools are primarily commercial and available for both small application contexts and large data integration application suites. In recent years, a market for data cleansing is also developing as a service.